\documentclass[prd,floatfix,preprintnumbers, superscriptaddress]{revtex4}
\usepackage{amsmath}
\usepackage{natbib}
\usepackage{graphicx,epsfig}
\setcitestyle{numbers,square,sort&compress}
\usepackage{url}

\newcommand{\bfl}{\begin{flushleft}}
\newcommand{\efl}{\end{flushleft}}
\newcommand{\bea}{\begin{eqnarray}}
\newcommand{\eea}{\end{eqnarray}}
\newcommand{\be}{\begin{equation}}
\newcommand{\ee}{\end{equation}}
\newcommand{\bi}{\begin{itemize}}
\newcommand{\ei}{\end{itemize}}

\def\bec{\begin{center}}
\def\eec{\end{center}}
\def\beq{\begin{equation}}
\def\eeq{\end{equation}}

\def\He{\theta}

\begin{document}

\title{Diffusion-limited Relic Particle Production}
\author{Robert J. Scherrer}
\affiliation{Department of Physics and Astronomy, Vanderbilt University,
Nashville, TN 37235, USA}
\author{Michael S. Turner}
\affiliation{Kavli Institute for Cosmological Physics, Chicago, IL  60637}
\affiliation{Department of Astronomy \& Astrophysics, University of Chicago,
Chicago, IL  60637}
\affiliation{Department of Physics, University of Chicago,
Chicago, IL  60637}
\begin{abstract}
We examine the thermal evolution of particle number densities in the early
universe when the particles have a finite diffusion length.  Assuming
that annihilations are impossible when the mean separation of the particles
is larger than their diffusion length, we derive a
version of the Boltzmann equation for freeze out in this scenario and an
approximate solution, accurate to better than 2\%.  The effect of a finite diffusion length
is to increase the final relic freeze-out abundance over its corresponding value when
diffusion effects are ignored.  When diffusion is limited only by scattering off of the thermal background,
and the annihilation cross section is bounded by unitarity, a significant effect
on the freeze-out abundance requires a scattering cross section much larger than the
annihilation cross section.  A similar effect is demonstrated when the relic particles
are produced via the freeze-in mechanism, but in this case the finite
diffusion length is due to
the scattering of particles that annihilate into the relic particle of interest.
For freeze in, the effect of a finite diffusion
length is to reduce the final relic particle abundance.  The effects
of a finite diffusion length are most important when the scattering cross
section or the relic mass are very large.  While we have not found a
particularly compelling example where this would affect previous results,
with the current interest in new dark matter candidates it could become
an important consideration.
\end{abstract}

\maketitle

\section{Introduction}

The evolution of relic particle densities is one of the central topics
in early-universe cosmology.  In the standard treatment, the relic particles are
taken to be initially relativistic and
in chemical and kinetic equilibrium with the
thermal background.  As the temperature $T$ drops below
the mass $m$ of these particles, they become nonrelativistic, and their
number density is Boltzmann suppressed as $e^{-m/T}$. Finally, the particles drop
out of thermal equilibrium with a fixed number density per comoving volume.
This freeze-out scenario has been
developed and refined over the past 50 years
\cite{zeldovich,chiu,LW,gs79,ST,KT,GK,GS,GG,SDB,BMS,Cannoni}. It remains
one of the favored models for the evolution of dark matter, and it can also
apply to the
evolution of other particles as well.

Here we consider an effect that has not been previously discussed in these
treatments of thermal evolution:  a finite diffusion length
for the evolving massive particles. 
In addition to undergoing
pair annihilations, these
particles will scatter off of the thermal background particles, resulting
in a finite diffusion length.  A finite diffusion length
could also result from more exotic scattering processes, such as interactions
with relic magnetic fields, domain walls, or other early-universe
phenomena. If this diffusion length is much larger
than the mean particle separation, then it has little effect on the particle
evolution.  However, if the
diffusion length drops below the mean particle separation during the freeze-out process, then
the effect on the evolution of the particle number density can be profound.

Although the effect discussed here has not been previously examined, there are
other discussions of annihilating, diffusing particles in the literature.
Zeldovich and Khlopov investigated the diffusion and annihilation of monopoles
in the early universe \cite{ZK}.  Several authors (see, e.g. Refs. \cite{TW,LW2})
examined systems of particles and antiparticles in a non-expanding background, with
initial inhomogeneities in the particle-antiparticle distribution, diffusion, and annihilation
(but no particle-antiparticle creation from the background).  These systems tend to evolve into
domains of particles and antiparticles within which further annihilation is impossible.

In the next section, we show how diffusion effects can be incorporated into the Boltzmann equation, and
we
derive an approximate solution to the diffusion-limited evolution of the relic particle number density that is accurate
to within 2\%.  In Sec. 3, we examine the particular case where scattering off of thermal background particles
limits the diffusion
length of the relic particles. In Sec. 4, we extend our discussion to the freeze-in
mechanism for relic particles. Our conclusions are summarized in Sec. 5.

\section{Diffusion and Freeze out}

Recall first the standard picture of thermal freeze out.  Consider a fermion $\chi$ with $g_\chi$ spin degrees of freedom.
For simplicity, we will take the particle to be its own antiparticle, but our results generalize easily to the case
of distiguishable particles and antiparticles, as well as to the case where $\chi$ is bosonic. 
The number density evolves as
\begin{equation}
\label{Boltzmann}
\frac{dn_\chi}{dt} + 3Hn_\chi = \langle\sigma_A v\rangle [n_{eq}^2 - n_\chi^2],
\end{equation}
where $n_\chi$ is the number density of the $\chi$ particles, $H$ is the Hubble parameter
($\equiv a^{-1}da/dt$, with $a$ being the scale factor), and $\langle \sigma_A v \rangle$ is the thermally-averaged cross section times relative velocity.
When $T \gg m$, the $\chi$ particles are highly relativistic, and the equilibrium number density $n_{eq}$ is given by
\begin{equation}
n_{eq} = (3/4)(g_\chi/\pi^2)\zeta(3) T^3,
\end{equation}
where $\zeta(3) \approx 1.202$.  (We take $\hbar = c = k =1$ throughout).
In the opposite limit, when $T \ll m$, we have instead
\begin{equation}
\label{nonrel}
n_{eq} = g_\chi \left(\frac{m_\chi T}{2 \pi} \right)^{3/2} \exp(-m_\chi/T).
\end{equation}
Following Refs. \cite{ST,KT}, we parametrize the annihilation cross section as
\begin{equation}
\langle \sigma_A v\rangle = \sigma_0 \left(\frac{m_\chi}{T}\right)^{-n},
\end{equation}
so that $n=0$ corresponds to $s$-wave annihilation, $n=1$ gives $p$-wave annihilation, and so on.
We make the standard change of variables  $x = m_\chi/T$ and $Y = n_\chi/s$, where $s$ is the entropy density,
given by
\begin{equation}
s = \frac{2 \pi^2}{45} g_* T^3,
\end{equation}
and $g_*$ is the  effective number of degrees of freedom in thermal equilibrium, all assumed to be
at the same temperature.
We obtain \cite{ST,KT}
\begin{equation}
\label{Yx}
\frac{dY}{dx} = - \lambda x^{-n-2}(Y^2 - Y_{eq}^2),
\end{equation}
where the constant $\lambda$ is given by
\begin{equation}
\label{lambda}
\lambda = 0.264 g_*^{1/2} m_{Pl} m_\chi \sigma_0,
\end{equation}
and $m_{Pl}$ is the
Planck mass.

In the standard scenario, $n_\chi$ tracks $n_{eq}$, and as $T$ drops below $m_\chi$ ($x > 1$), the $\chi$
number density becomes exponentially suppressed as in Eq. (\ref{nonrel}).  When $x$ reaches a value
of $x_f \sim 10-30$, the rates for the reactions that produce $\chi$ become negligible compared to the annihilation
rate, and the abundance freezes out.  However, annihilations continue, reducing the value of $Y$ relative
to its value at $x = x_f$.  The final abundance is well-approximated by \cite{ST,KT}
\begin{equation}
\label{Yinf}
Y_\infty = \frac{n+1}{\lambda}x_f^{n+1}.
\end{equation}

This derivation assumes an effectively infinite mean free path for the particles.  However,
in addition to annihilations, the $\chi$ particles scatter off of the thermal background, resulting
in a finite diffusion length $d$, which will be a function of $T$.
(Here we are taking $d$ to be the physical, not
the comoving diffusion length).
Although scattering is the only unavoidable source of a finite diffusion length, it is not
the only possibility.  Magnetic fields, domain walls, or other early-universe phenomena might
also reduce the $\chi$ diffusion length.  Hence, we will take $d(T)$ for now to be a free
parameter and determine under what conditions a finite diffusion length affects the freeze-out process.  In
the next section, we will examine the specific case of scattering interactions with thermal background particles.

Now consider how a finite value of $d$ alters the freeze-out process.  It has no effect at all on the creation
of $\chi$ particles, since the creation
rate is determined by the rate of annihilation of thermal background particles into $\chi$ particles.
However, a finite diffusion length {\it does} affect the $\chi \chi$ annihilation rate.  A given $\chi$
particle can only travel a distance $d(T)$ at a temperature $T$ to annihilate with another $\chi$ particle.
Hence, when $n_\chi \ll 1/d^3$, the annihilations effectively cease.  Conversely, when $n_\chi \gg 1/d^3$,
diffusion has no effect on $\chi \chi$ annihilations.  We will make the approximation that annihilation is completely
unaffected for $n_\chi > 1/d^3$ and that annilations cease completely for
$n_\chi < 1/d^3$.  In reality,
this behavior will not be a step function but will vary more gradually with $n_\chi$ and $d$.  However,
our approximation, while admittedly crude, will be sufficient for a first calculation of the effect
considered here.

With this approximation, we can multiply $n_\chi^2$ in Eq. (\ref{Boltzmann}) by the appropriate step function
$\He(n - 1/d^3)$, where the Heaviside step function $\He(x)$ is defined by $\He(x) = 0$ for $x<0$ and $\He(x) = 1$ for
$x >0$.  This gives
\begin{equation}
\frac{dn_\chi}{dt} + 3Hn_\chi = \langle\sigma_A v\rangle[n_{eq}^2 - n_\chi^2 \He(n_\chi - 1/d^3)].
\end{equation}
Consider how this change alters the evolution of the particle abundances.  As long as $n_\chi > 1/d^3$,
the evolution is unaffected, but when $n_\chi$ drops below $1/d^3$, the annihilations shut off.  However,
creation of $\chi$ continues, ultimately pushing $n_\chi$ back above $1/d^3$.  The net affect is
that in thermal equilibrium, $n_\chi$ does not necessarily track $n_{eq}$; instead, it tracks the {\it larger} of
$n_{eq}$ and $1/d^3$.  Effectively, $1/d^3$ gives a floor abundance on $n_\chi$.

To determine how this alters the freeze-out abundance, we make the same change of variables as in 
Eq. (\ref{Yx}), so that
\begin{equation}
\label{Ydif}
\frac{dY}{dx} = -\lambda x^{-n-2}[Y^2 \He(Y - Y_d) - Y_{eq}^2],
\end{equation}
where we have defined the new quantity $Y_d$ to be given by
\begin{equation}
\label{Yddef}
Y_d = (1/d^3)/s.
\end{equation}
The quantity $Y_d$ has a simple physical interpretation; it is the inverse of the total entropy in
a diffusion volume $d^3$.

To keep our argument as general as possible, we let $d$ be an arbitrary power-law in $T$, namely
\begin{equation}
d(T) = d_0 (T/T_0)^{-\alpha},
\end{equation}
where $d_0$ and $T_0$ are arbitrary fiducial values of the diffusion length and the temperature,
and we are implicitly assuming that the diffusion length increases as the temperature decreases ($\alpha > 0$).
Because
$s$ scales as $a^{-3} \sim T^3$, we can then write $Y_d$ as
\begin{equation}
\label{YdY1}
Y_d = Y_1 x^{3-3\alpha},
\end{equation}
where the fiducial quantity $Y_1$ is defined by Eq. (\ref{YdY1}): $Y_1$ is the value of $Y_d$ at $x = 1$.  Hence,
$Y_1$ is the inverse of the entropy in a diffusion volume $d^3$ at a temperature $T = m_\chi$.
Larger $Y_1$ corresponds to smaller diffusion length and vice-versa.

\begin{figure}[tbh]
\centerline{\epsfxsize=6truein\epsffile{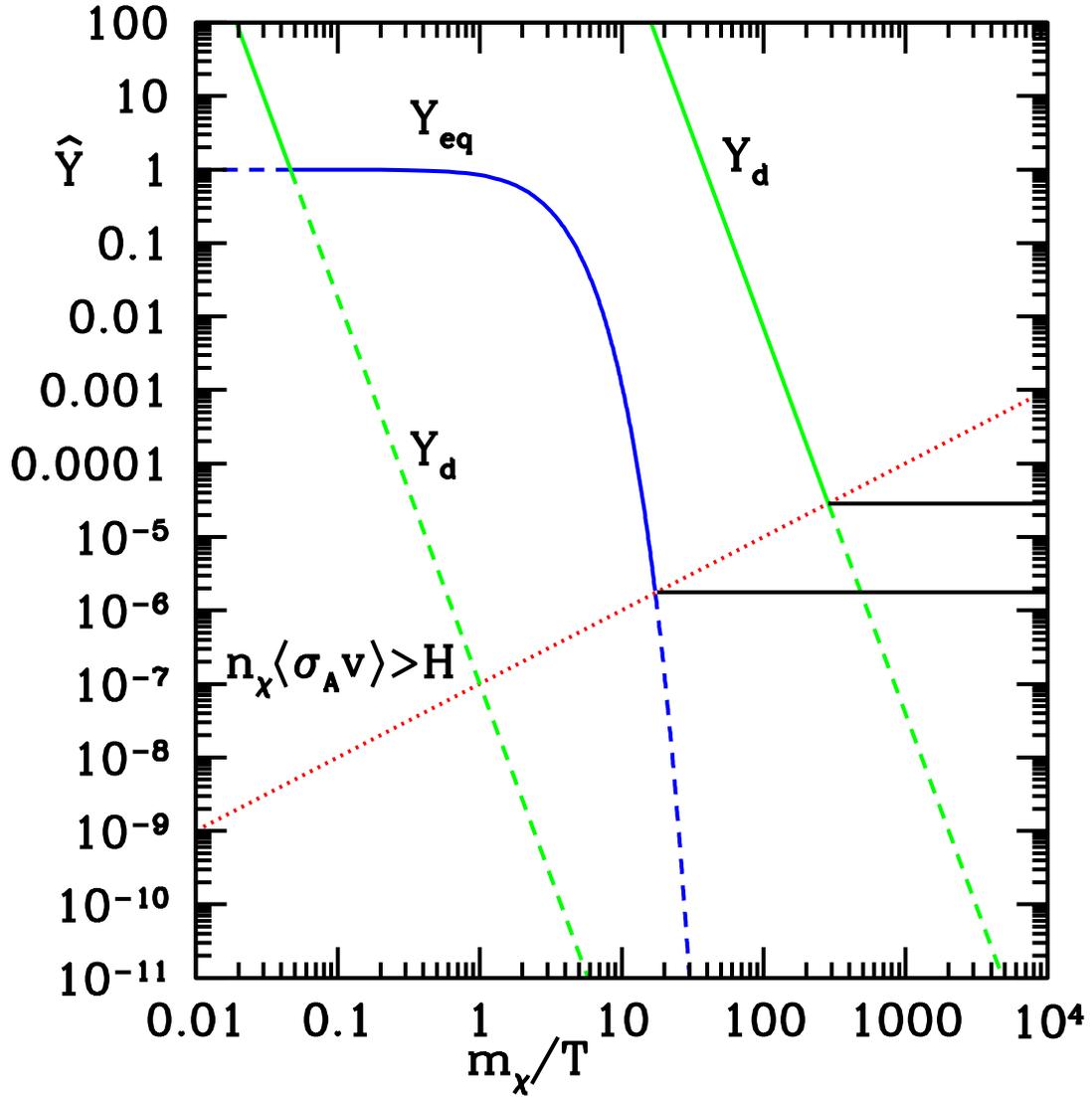}}
\caption{A rough sketch of the evolution of $\hat Y$, the
value of $Y$ normalized
to its relativistic value, $n_{eq}/s$, as a function
of $m_\chi/T$, for the case of $s$-wave annihilations.
(Here $\hat Y \equiv Y/(n_{eq}/s)$, with $n_{eq}/s = 0.208 g_\chi/g_*$ for a
relativistic fermion).
Blue curve is the thermal equilibrium abundance $Y_{eq}$.  Area above
the dotted red line is the region in which $n_\chi \langle \sigma_A v \rangle > H$, so annihilations can take place.  (We
have fixed
$\langle \sigma_A v \rangle$ to a single illustrative value).
Green lines give $Y_d$ as defined in Eq. (\ref{YdY1}) for $\alpha = 11/4$ and two different values of $Y_1$.
For each value of $Y_1$, the value of $\hat Y$ tracks the larger of $Y_{eq}$ and
$Y_d$ until $\hat Y$ drops
below the threshold for annihilations given by the dotted red line, at which point the abundances
freeze out at the values shown by the horizontal black lines.  In both cases,
the value of $\hat Y$ traces out the solid portion of the displayed curves.}
\end{figure}

A rough approximation to the evolution of $Y$ is sketched out in Fig. 1 for the case of $s$-wave annihilation.
For illustrative purposes, we take $\alpha = 11/4$ here and in Fig. 2, as this corresponds
to a diffusion length limited by scattering off of thermal background particles with a constant
scattering cross section (see the next section).
However, our qualitative results do not
depend on the value of $\alpha$.
When $\chi$ is in thermal equilibrium, with $Y \approx Y_{eq}$,
a finite diffusion length has no effect on the evolution of $Y$ as
long as $Y_{eq} > Y_d$.  However, when $Y_{eq} < Y_d$, the value of $Y$
tracks $Y_d$ instead of $Y_{eq}$.  Thus, $Y$ tracks the {\it larger}
of $Y_{eq}$ or $Y_d$, until $n_\chi \langle \sigma_A v \rangle$ drops below $H$ and the abundance freezes
out at the value shown in Fig. 1 by the horizontal black lines.
(Note that if $Y_d > Y_{eq}$ when
$\chi$ is highly relativistic, we will have $Y > Y_{eq}$ and $n_\chi > n_{eq}$. This is a rare case
in which it is possible for the density of a particle to exceed its relativistic equilibrium density.  Physically,
this occurs because the finite diffusion length prevents particle annihilations, while production from
the thermal background continues to produce $\chi$ particles.  This would require an exceedingly small diffusion length).

This approximation assumes a sudden sharp freeze out, but in reality residual
annihilations continue to occur even after $n_\chi \langle \sigma_A v \rangle$ drops below $H$.  This is illustrated in
Fig. 2, in which we display
the exact evolution, derived from a numerical integration of Eq. (\ref{Ydif}), for $s$-wave annihilation.
\begin{figure}[tbh]
\centerline{\epsfxsize=6truein\epsffile{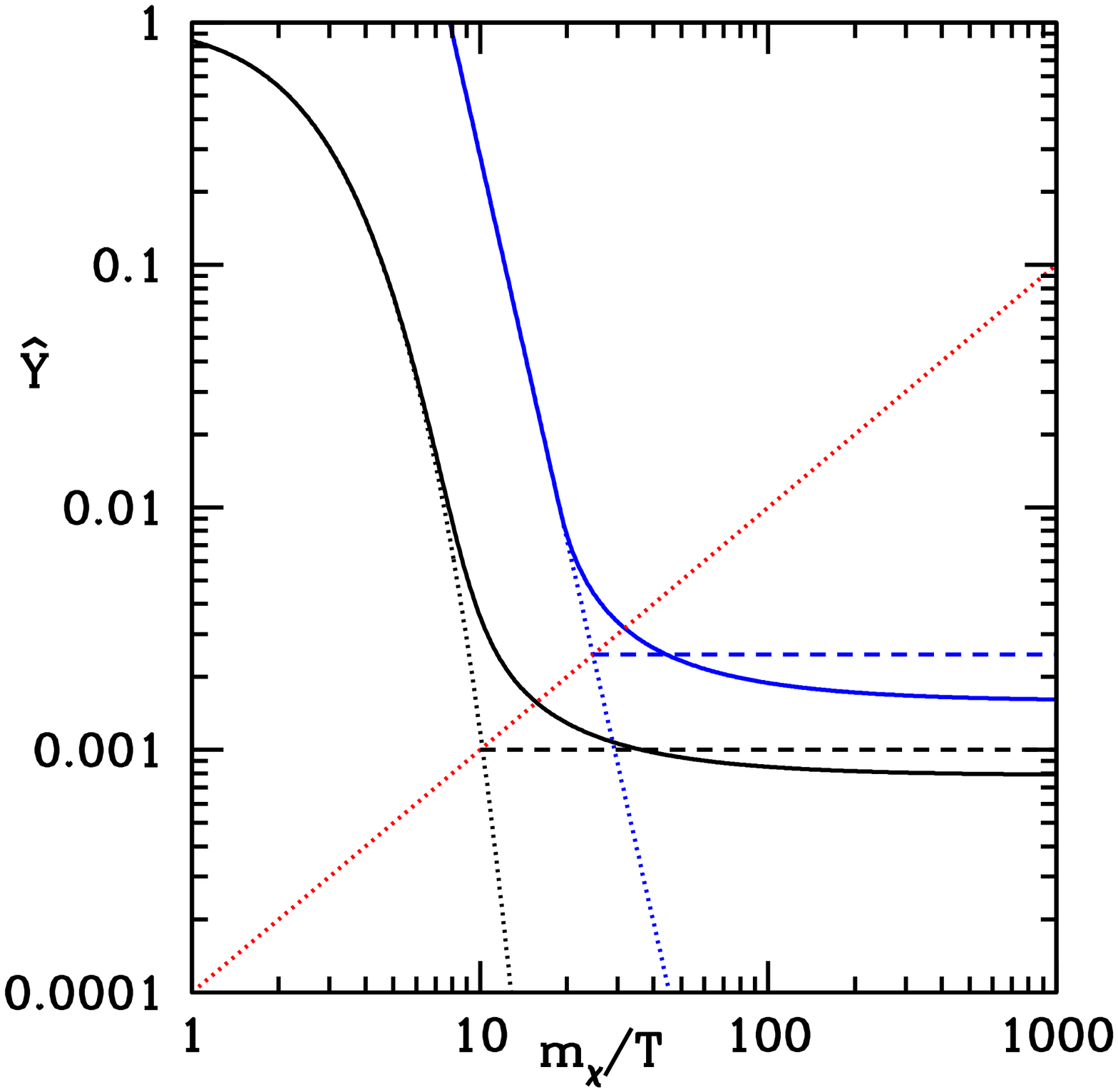}}
\caption{The evolution of $\hat Y$, defined as in Fig. 1, as a function
of $m_\chi/T$,
from a numerical integration of Eq. (\ref{Ydif}), for $s$-wave annihilations with
$\lambda (n_{eq}/s) = 10^4$ and $\alpha = 11/4$.
Solid black curve gives the evolution in the standard model
with no diffusion effects.  Dotted black curve is $Y_{eq}/(n_{eq}/s)$.  Solid blue curve
gives evolution
for $Y_1/(n_{eq}/s) = 50000$; dotted blue line gives the corresponding
value of $Y_d/(n_{eq}/s)$.  Dotted red line corresponds to
$n_\chi \langle \sigma_A v \rangle = H$.
The points at which this line intersects the dotted blue
line and the dotted black line correspond to
the sudden freeze-out approximation in the diffusion-limited case and in the standard model, respectively, with the resulting
abundances given by the dashed blue and dashed black lines.}
\end{figure}
The solid blue curve in Fig. 2 corresponds to the exact evolution
where $Y_{eq} < Y_d$ at $x = x_f$, where $x_f$ is the
freeze-out value of $x$ in the standard scenario without diffusion effects.
The dashed blue curve shows the equivalent abundance in the sudden freeze-out
approximation.
As in the case of standard freeze out, residual annihilations reduce the final
abundance below that obtained by using the sudden freeze out approximation.

We now calculate an expression for the final abundance that takes into account
these residual annihilations.  For the diffusion-limited case,
we have $Y \approx Y_d$ instead of $Y \approx Y_{eq}$ at $x = x_f$.
The reactions that create $\chi$ particles effectively shut off at $x_f$,
and Eq. (\ref{Ydif}) becomes
\begin{equation}
\label{Ydecdif}
\frac{dY}{dx} = -\lambda x^{-n-2} Y^2 \He(Y - Y_d),
\end{equation}
At this point, $Y$ continues to track $Y_d$, which is a decreasing function
of $x$ given Eq. (\ref{YdY1}).  The rate at which $Y_d$ decreases with $x$
is then
\begin{equation}
\label{Ydd}
\frac{dY_d}{dx} = Y_1(3-3\alpha)x^{2-3\alpha}. 
\end{equation}
However, $Y$ cannot decrease faster than the annihilation
rate given by Eq. (\ref{Ydecdif})
in the absence of diffusion effects, namely
\begin{equation}
\label{Ydec}
\frac{dY}{dx} = -\lambda x^{-n-2} Y^2.
\end{equation}
Hence, $Y$ tracks $Y_d$ only until $|dY_d/dx|$ from Eq. ({\ref{Ydd})
becomes larger than $|dY/dx|$ from Eq. (\ref{Ydec}).  At this point,
diffusion effects become irrelevant and the further evolution of
$Y$ is determined only by Eq. (\ref{Ydec}), as in standard freeze out.

We can now calculate the final abundance of $Y$ in the presence
of diffusion effects, which we will denote $\widetilde Y_\infty$.
Equating $dY_d/dx$ from Eq. ({\ref{Ydd}) with $dY/dx$ from Eq. ({\ref{Ydec})
and taking $Y = Y_d$ given by Eq. ({\ref{YdY1}) gives the value of $x$ at which
annihilations can proceed without diffusion effects, which we will denote $x_d$:
\begin{equation}
x_d = \left(\frac{\lambda Y_1}{3 \alpha -3}\right)^{1/(3\alpha-2+n)}.
\end{equation}
At $x = x_d$, the value of $Y$ is given by the value of $Y_d$ in Eq. (\ref{YdY1}), namely
\begin{equation}
Y(x = x_d) = Y_1 \left(\frac{3 \alpha -3}{\lambda Y_1}\right)^{(3\alpha-3)/(3\alpha-2+n)}
\end{equation}
In order to calculate the new asymptotic value of $Y$, denoted by $\widetilde Y_\infty$, we integrate
Eq. (\ref{Ydec}) from $x = x_d$ to $x = \infty$, using the value of $Y(x = x_d)$ that we have just derived.
The final result is
\begin{equation}
\label{Yinfn2}
\widetilde Y_\infty = \left(\frac{(3 \alpha - 3)(1+n)}{3 \alpha -2+n}\right)\frac{1}{\lambda}
\left[\frac{\lambda Y_1}{3 \alpha -3} \right]^{(1+n)/(3\alpha - 2 + n)}.
\end{equation}
Alternately, we can express
the value of $\widetilde Y_\infty$ relative to $Y_\infty$,
the asymptotic value of $Y$ for the case where diffusion effects
are negligible, which is given by
Eq. (\ref{Yinf}).  This gives
\begin{equation}
\label{Yinfn3}
\frac{\widetilde Y_\infty}{Y_\infty} = \left(\frac{3 \alpha - 3}{3 \alpha -2+n}\right)\frac{1}{x_f^{n+1}}
\left[\frac{\lambda Y_1}{3 \alpha -3} \right]^{(1+n)/(3\alpha - 2 + n)}.
\end{equation}
Eqs. (\ref{Yinfn2})-(\ref{Yinfn3}) are the main results of our paper.
They take a particularly simple form for $s$-wave annihilations ($n=0$), namely
\begin{equation}
\label{Yinf2}
\widetilde Y_\infty = \left(\frac{3 \alpha - 3}{3 \alpha -2}\right)\frac{1}{\lambda}
\left[\frac{\lambda Y_1}{3 \alpha -3} \right]^{1/(3\alpha - 2)},
\end{equation}
and
\begin{equation}
\label{Yinf3}
\frac{\widetilde Y_\infty}{Y_\infty} = \left(\frac{3 \alpha - 3}{3 \alpha -2}\right)\frac{1}{x_f}
\left[\frac{\lambda Y_1}{3 \alpha -3}\ \right]^{1/(3\alpha - 2)}.
\end{equation}
Comparing to a numerical integration
of Eq. (\ref{Ydif}),
we find that these expressions for $\widetilde Y_\infty$ are accurate to better than 2\% for both the
$s$-wave and $p$-wave cases.

\section{Finite diffusion length from scattering}

Our results in the previous section apply to the general case
in which the diffusion length of the particles is small enough to
affect the freeze-out process, without reference to any specific model
of diffusion.  Here we will consider the specific case of scattering
of the $\chi$ particles off of standard-model thermal background particles.
This process has been explored in detail in connection with the process of
kinetic decoupling, which generally occurs later than chemical decoupling for
most particle species \cite{Chen,BH,Dent,VG,BIKW,Gondolo,WEI,Kamada}.  Here we will be interested in the extreme case
for which the scattering rate is large enough to affect the freeze-out process itself,
as discussed in the previous section.

Assume that the $\chi$ particles can annihilate into thermal Standard Model (SM)
particles, and also scatter off of those same particles:
\begin{equation}
\chi + \chi \rightarrow SM + SM,
\end{equation}
and
\begin{equation}
\chi + SM \rightarrow \chi + SM.
\end{equation}
We will assume $s$-wave annihilations, so that
$\langle \sigma_A v \rangle = \sigma_0$, while the cross section for scattering off of standard model
particles will be $\sigma_S$.

As the $\chi$ particles scatter off of the thermal background particles, they undergo a random walk
with step size $l$, which gives a diffusion length $d$:
\begin{equation}
d = \sqrt{l v t}.
\end{equation}
When $T > m_\chi$, the value of $l$ is just the mean free path, given by $l =
(n_S \sigma_S)^{-1}$.
However, for $T < m_\chi$, multiple scatterings are required to significantly change
the momentum of the $\chi$ particles.  Following Ref. \cite
{VG}, we will assume that the
number of such collisions required to alter the trajectory of a given $\chi$ is $\sim m_\chi/T$.
Thus, the step size $l$ in this case is $l \sim (n_S \sigma_S)^{-1} (m_\chi/T)$.  Since freeze out occurs
when $m_\chi /T > 1$, we will use the latter expression to derive the diffusion length.  Then we obtain
\begin{equation}
\label{d}
d = \sqrt {(n_S \sigma_S)^{-1}(m_\chi /T) vt}.
\end{equation}
We take $v \approx \sqrt{3T/m}$ and
$t = 0.3 g_*^{-1/2}m_{Pl}/T^2$, which assumes a radiation-dominated universe.
The values of $n_S$ and $\sigma_S$ are necessarily model dependent.
To make an estimate of the circumstances under which scattering can
alter the freeze-out process, we will assume that $\chi$ can scatter off
of all of the degrees of freedom in the relativistic background, so that
$n_S$ is given by $n_S = \zeta(3) g_{n} T^3/\pi^2$, where $g_{n} = 3/4 ~( =1)$ per fermionic
(bosonic)
degree of freedom.
Because the scattering
cross section itself is completely model dependent, we will, for simplicity,
take $\sigma_S$ to be a constant.  (For a discussion of other possibilities,
see, e.g., Ref. \cite{Chen}).
It is straightforward to generalize our results to other functional
forms for $\sigma_S$.

Combining these expressions, and taking $g_n \approx g_*$, we obtain
\begin{equation}
d = 2.1 g_*^{-3/4} \sigma_S^{-1/2} m_{Pl}^{1/2} m_\chi^{1/4} T^{-11/4}.
\end{equation}
Using the standard expression for entropy density, we can substitute this value for
the diffusion length into Eq. (\ref{Yddef}) to derive an expression for $Y_d$:
\begin{equation}
Y_d = 0.26 g_*^{5/4} \sigma_S^{3/2} m_{Pl}^{-3/2} m_\chi^{9/2}
\left(\frac{m_\chi}{T}  \right)^{-21/4}.
\end{equation}
In terms of the parameters of Eqs. (\ref{YdY1}) and (\ref{Yinf2}), we then have:
\begin{equation}
\label{Y1scat}
Y_1 = 0.26 g_*^{5/4} \sigma_S^{3/2} m_{Pl}^{-3/2} m_\chi^{9/2},
\end{equation}
and
\begin{equation}
\label{alphascat}
\alpha = 11/4.
\end{equation}

Using the values for $Y_1$ and $\alpha$ from Eqs. (\ref{Y1scat}) and
(\ref{alphascat}), along with the definition of $\lambda$
from Eq. (\ref{lambda}), Eq. (\ref{Yinf3})
becomes
\begin{equation}
\frac{\widetilde Y_\infty}{Y_\infty} = \frac{21}{25} \frac{1}{x_f} \left[0.013
g_*^{7/4} m_{Pl}^{-1/2} m_\chi^{11/2} \sigma_0 \sigma_S^{3/2}
\right]^{4/25}.
\end{equation}
Clearly, the change in the relic $\chi$ abundance relative to its
standard abundance is an increasing function of the annihilation
cross section, the scattering cross section, and the particle
mass.
Diffusion effects become important when $\widetilde Y_\infty/Y_\infty > 1$, which
corresponds to
\begin{equation}
\label{masslimit}
m_\chi > 2.7 ~x_f^{25/22}~ g_*^{-7/22} \sigma_0 ^{-2/11} \sigma_s^{-3/11} m_{Pl}^{1/11}.
\end{equation}
Taking $g_* \sim 100$ and $x_f \sim 10$ (both good order-of-magnitude
approximations), we graph the region for which scattering affects the freeze-out abundance
in Fig. 3.
\begin{figure}[tbh]
\centerline{\epsfxsize=6truein\epsffile{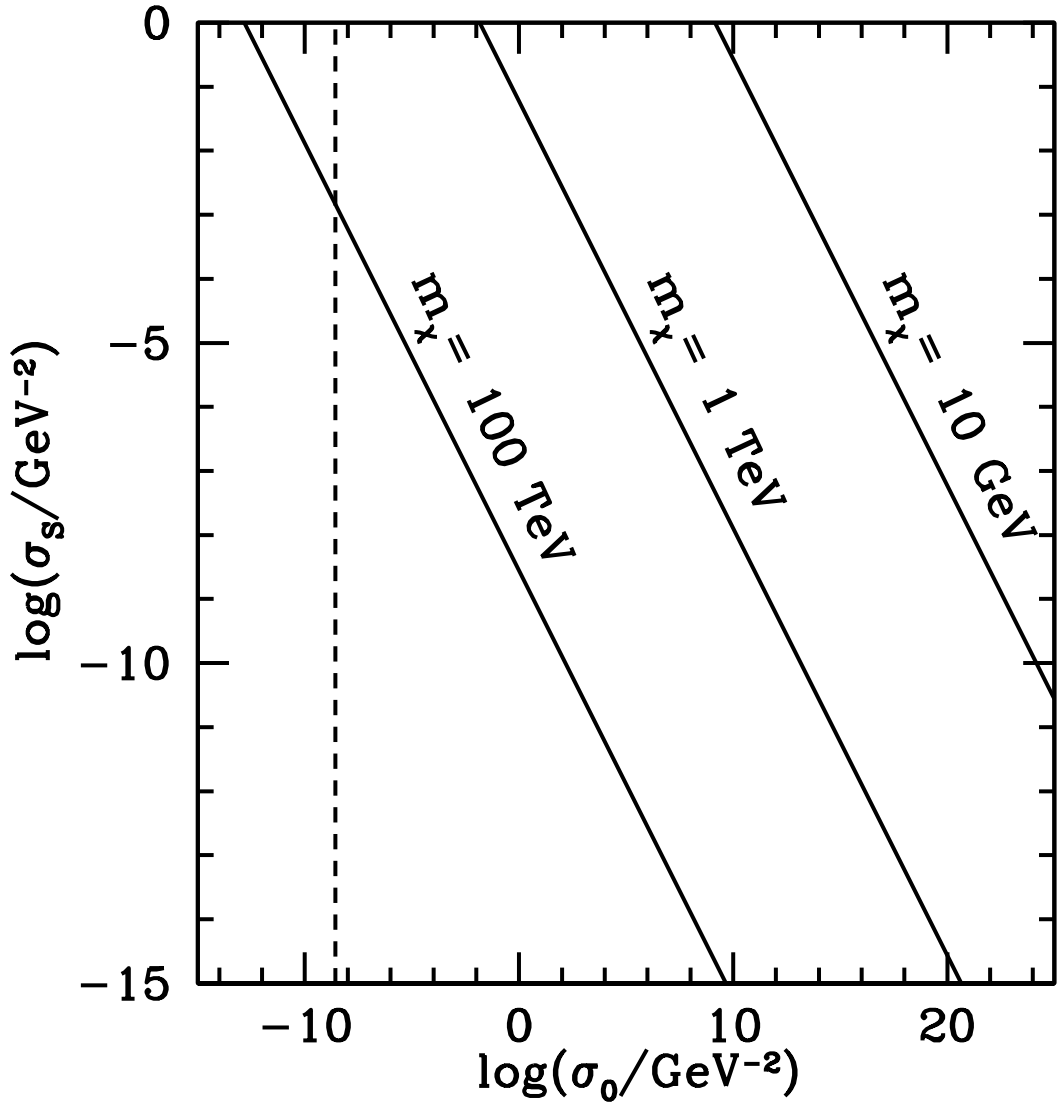}}
\caption{Region in parameter space defined by the particle mass
$m_\chi$, $s$-wave annihilation cross section $\sigma_0 \equiv \langle \sigma_A v
\rangle$, and scattering cross section $\sigma_S$, for which a finite
diffusion length from scattering off of the thermal
background will alter the standard freeze-out abundance
of $\chi$:  this region lies
above the solid lines corresponding to each indicated mass.  Vertical
dashed line is the annihilation cross section corresponding
to the observed dark matter abundance today in the absence
of diffusion effects ($\sigma_0 = 2.6 \times 10^{-9}$ GeV).}
\end{figure}
In this figure, the area above the solid line corresponding to each
indicated mass represents the region in parameter space for which a
finite diffusion length from scattering
will increase the relic freeze-out abundance relative
to its value in the absence of diffusion effects.  For reference,
we also include the value of $\sigma_0$ that corresponds to
the observed dark matter abundance for a thermal relic,
$\sigma_0 \sim 1.0 \times 10^{-36}$ cm$^2
= 2.6 \times 10^{-9}$ GeV$^{-2}$.

Note also that $m_\chi$ is bounded from above by
unitarity, which requires that \cite{GK}
\begin{equation}
\label{unitarity}
\sigma_0 m_\chi^2 < 4 \pi.
\end{equation}
This bound requires particles with masses above roughly 100 TeV to have an annihilation cross section
below the standard weak-scale $\sigma_0$ cited above, which results in a freeze-out abundance
larger than the observed dark matter abundance.  (See, however, Ref. \cite{Unwin} for mechanisms to
evade this bound).  For particle masses below the unitarity bound and $\sigma_0 = 2.6 \times 10^{-9}$ GeV$^{-2}$,
we see from Fig. 3 that the scattering cross section must be much larger than the annihilation cross section
in order for a finite diffusion length to alter the freeze-out abundance.

\section{Diffusion-limited Freeze In}

Another mechanism to produce relic particles, called freeze in, was first proposed by Hall et al. \cite{Hall}.
In this scenario, a feebly interacting massive particle (FIMP) is coupled
too weakly to standard model particles to ever be in equilibrium with the thermal
background.  However the FIMP can be produced by annihilations of other particles,
albeit at a very low rate.  Ref. \cite{Hall} gives a number
of variations on this theme involving a combination of freeze in, freeze out, and particle decays, but
we will confine our attention here to the simplest scenario, in which there is only freeze in of a relic particle.
Because of the variety of possible scenarios, we will keep our discussion as general as possible.  Nonetheless, we will
be able to derive some interesting results for freeze in.

In this simple scenario, we assume that the particle $\chi$ is produced through annihilation of one or more
particles $A$, so that the production rate is
\begin{equation}
\label{freezeeq}
\frac{dn_\chi}{dt} + 3Hn_\chi = \langle \sigma_A v \rangle n_A^2.
\end{equation}
Again, we define $Y_\chi$ and $Y_A$ to be the ratio of the particle number densities to the entropy
density, so Eq. (\ref{freezeeq}) becomes
\begin{equation}
\label{freezeineq2}
\frac{dY_\chi}{dt} = s \langle \sigma_A v \rangle Y_A^2.
\end{equation}
In the treatment of Ref. \cite{Hall}, it is assumed that $A$ is a standard-model particle in equilibrium
with the thermal background,
so that
when $T$ drops below $m_\chi$, the interactions producing the
$\chi$ are Boltzmann suppressed and freeze in terminates.  To keep our results as general as possible,
we will not make these assumptions, but we will take
$A$ to be relativistic in our epoch of interest, as this simplifies the calculation.

Now consider the effect of a finite diffusion length on the freeze-in process.
In this case, the diffusion length of $\chi$ is irrelevant, since it is
the annihilation of $A$ that produces the final $\chi$ abundance, so it is the finite diffusion length of $A$ that has
the potential to alter this abundance.  As before, let $d$ be the diffusion length of $A$.  Then
the freeze in process will be altered whenever $n_A d^3 < 1$.

Now consider the case where the finite diffusion length is produced by the scattering of $A$ off of particles
in the thermal background.
Since we are taking $A$ to be relativistic at $T \sim m_\chi$, Eq. (\ref{d}) becomes
\begin{equation}
d = \sqrt{(n_S\sigma_S)^{-1} t},
\end{equation}
and the condition for the finite diffusion length to alter the relic abundance, $n_A d^3 < 1$, becomes
\begin{equation}
n_A n_S^{-3/2} \sigma_S^{-3/2} t^{3/2} < 1.
\end{equation}
We now take $n_S \sim T^3$, $n_A  \sim Y_A T^3$, and
$t \sim m_{Pl}/T^2$, to obtain
\begin{equation}
\label{Tfreeze}
Y_A T^{-9/2} m_{Pl}^{3/2} \sigma_S^{-3/2} < 1.
\end{equation}
Equation (\ref{Tfreeze}) is our condition for which a finite diffusion length for $A$
alters the freeze-in process at temperature $T$,
due to scattering of $A$ off of the thermal background.
In the case of freeze in, finite diffusion
effects suppress the annihilation of $A$ into $\chi$ particles, reducing the final freeze-in abundance of $\chi$.
The size of this effect, and whether it occurs at all, depend upon the detailed scenario for freeze in.

\section{Conclusions}

It is remarkable that after 50 years, we are still finding new aspects
of the thermal evolution of relic particle abundances.  It is important
to note that the effects we have outlined here are not ``optional."
Any particle undergoing annihilation into standard model particles
will also scatter off of the thermal background, and the only
question is the magnitude of this effect.  For particles satisfying the
unitarity bound, scattering effects
will affect the freeze-out abundance only if 
the scattering cross
section is much larger than the annihilation cross section.  However,
the scattering effects outlined in the previous section are unlikely
to exhaust the possibilities of diffusion-limited freeze out; one
can also consider scattering from relic magnetic fields, domain walls, or other
exotic early-universe phenomena.

The main approximation we made in our derivation of the effect of a finite diffusion
length is the sharp cut-off in the annihilation rate when the mean particle separation
is larger than the diffusion length.  A more detailed calculation would show a more
gradual effect.  However, we expect the derivation in Sec. 2 to be qualitatively accurate
and, furthermore, independent of the particular mechanism limiting the particle diffusion.
In this derivation we have assumed nothing about the actual mechanism producing a finite
diffusion length; our result depends only the value of the diffusion length as a function of
temperature.

Our calculation in Sec. 3 is more model-dependent; the effect of scattering will depend
on the total scattering cross section, the particles off of which scattering occurs, and the
scaling of scattering with temperature.  However, the formalism we have developed in that
section is easily extended to other scattering scenarios such as those we have mentioned above.

A finite diffusion length from scattering can also affect the freeze-in process
for relic particle production.  In this case, the effects depend not on the scattering
of the relic particle itself, but on the scattering of the particles that annihilate into the relic particle.
For freeze in, the effect is the opposite of the effect on freeze out: a finite diffusion length
decreases the relic particle abundance.

While we have not found an especially compelling example for which a finite diffusion length would significantly
alter previous results for freeze out or freeze in, the current heightened interest in new dark matter candidates
suggests the possibility that diffusion could be an important consideration in computing relic abundances in the
future.  We can say more generally that the effects discussed here are likely to be most important when the
relic particle mass or the scattering cross section is very large.

\section*{Acknowledgments}
R.J.S. was supported in part by the Department of Energy
(DE-SC0019207).  M.S.T. was supported at the University of Chicago by
the Kavli Institute for Cosmological Physics through grant NSF PHY-1125897
and an endowment from the Kavli Foundation and its founder Fred Kavli.

\end{document}